\documentclass[useAMS,usenatbib]{mn2e}

\usepackage{epsf}
\usepackage{graphicx}

\def\ltsim{\, {}^<_\sim \,}
\def\vv{\hbox{\em V\/}}
\def\bb{\hbox{\em B\/}}
\def\bmd{\hbox{\em B--DDO$\,$51\/}}
\def\dmv{\hbox{\em DDO$\,$51--V\/}}
\def\bmv{\hbox{\em B--V\/}}
\def\bmi{\hbox{\em B--I\/}}
\def\vmi{\hbox{\em V--I\/}}

\def\umb{\hbox{\em U--B\/}}
\def\umd{\hbox{\em U--DDO$\,$51\/}}
\def\sec{${}^{\prime\prime}$}

\newcommand{\feh}{\hbox{$ [{\rm Fe}/{\rm H}]$ }}

\title[Formation of the Globular Cluster IC$\,$4499]{Constraints on the Formation of the Globular Cluster IC$\,$4499 from Multi-Wavelength Photometry\thanks{Based in part on observations 
made with the European Southern Observatory
telescopes obtained from the ESO/ST-ECF Science Archive Facility.}\\}
\author[A. R. Walker et~al.]{A. R. Walker$^{1}$\thanks{E-mail:
awalker@ctio.noao.edu}, 
A. M. Kunder$^{1}$, G. Andreuzzi$^{2}$, A. Di Cecco$^{3}$, P.B. Stetson$^{4}$,
\newauthor
M.Monelli$^{5}$, S. Cassisi$^{6}$, G. Bono$^{3}$, R. De Propris$^{1}$, 
M. Dall'Ora$^{7}$, J. M. Nemec$^{8}$, 
\newauthor
M. Zoccali$^{9}$
\\ 
$^{1}$Cerro Tololo Inter-American Observatory, National Optical Astronomy Observatory, Casilla 603, La Serena, Chile\\
$^{2}$Fundaci\'{o}n Galileo Galilei - INAF, Bre\~{n}a Baja, La Palma, Spain \\
$^{3}$Departimento di Fisica, Universit\'{a} di Roma Tor Vergata, via della Ricerca Scientifica 1, 00133, Rome, Italy \\
$^{4}$Dominion Astrophysical Observatory, Herzberg Institute of Astrophysics, National Research Council,Victoria, British Columbia V9E 2E7, Canada \\
$^{5}$IAC, Calle Via Lactea, E38200, La Laguna, Tenerife, Spain and Departamento de Astrof\'{i}sica, Universidad de La
Laguna, Tenerife, Spain \\
$^{6}$INAF-Osservatorio Astronomica di Collurania, via M. Maggini, 64100 Teramo, Italy \\
$^{7}$INAF-Osservatorio Astronomico di Capodimonte, via Moiarello 16, 80131, Naples, Italy \\
$^{8}$Department of Physics \& Astronomy, Camosun College, Victoria, British Columbia V8P 5J2, Canada \\
$^{9}$Depto de Astronomia y Astrofisica, P. Universidad Catolica, Casilla 306, Santiago 22, Chile \\
}
\begin{document}

\date{accepted for publication, MNRAS, Mar 18, 2011}

\pagerange{\pageref{firstpage}--\pageref{lastpage}} \pubyear{2002}

\maketitle

\label{firstpage}

\begin{abstract}

We present new multiband photometry for the Galactic globular cluster IC$\,$4499 
extending well past the main sequence turn-off in the $U$, $B$, $V$, 
$R$, $I$, and $DDO$\,$51$ bands.  This photometry is used to determine 
that IC$\,$4499 has an age of 12$\pm$1 Gyr and a cluster
reddening of $E(B-V) =$0.22$\pm$ 0.02.  Hence, IC$\,$4499 is 
coeval with the majority of Galactic GCs, in contrast to suggestions of a 
younger age.  The density profile of the cluster is observed to not flatten out 
to at least $r\sim$800\sec, implying that either the tidal radius of this
cluster is larger than previously estimated, or that IC$\,$4499 is surrounded
by a halo.  Unlike the situation in some other, more massive, globular clusters,
no anomalous color spreads in the UV are detected among the red giant
branch stars.  The small uncertainties in our photometry should allow the 
detection of such signatures apparently associated with variations of
light elements within the cluster, suggesting that IC$\,$4499 consists 
of a single stellar population.
\end{abstract}


\section{Introduction}

Galactic globular clusters (GCs) are populous collections of old stars dating from the 
first few Gyr after the Big Bang.  Because they typically consist of $10^5$ to $10^6$ stars, 
even short-lived evolutionary phases can be well populated.  The wide range in metal abundances,
as well as their extensive lifetimes, make globular clusters key ingredients in any 
explanation of the stellar and chemical evolution of our Galaxy.  With few exceptions, GCs 
have long been thought to be excellent examples of single stellar populations
\citep[SSPs,][]{ren86} that formed at about the same time as our Galaxy.  However, many GCs show
light element abundance spreads indicative of the presence of multiple (but nearly coeval)
populations.

The main goal of this paper is to take a fresh look 
at IC$\,$4499, a Galactic GC with a mass somewhat lower than those clusters that most 
often clearly exhibit complex photometric behavior.  If lower mass 
clusters correspond more closely to single stellar populations than the 
highest mass clusters, then 
the interpretation of their presumably simpler histories may be critical in deciphering 
the early evolution of our Galaxy and its system of GCs.  Although major 
HST programs such as \citet{sar07} are providing outstanding data sets for the 
nearby clusters, many of the lower mass 
clusters ($\sim 5$ x $10^4 M_{\odot}$) still lack definitive study.   

The very lowest mass clusters are not tractable to study as they do not
contain statistically significant numbers of stars in the various
evolutionary phases.   However, moderate mass clusters such as 
IC$\,$4499 do contain sufficient stars for population analyses and are relatively
numerous in our Galaxy.  Any explanation of their properties should be
consistent with the exotic behavior seen in the generally more massive
clusters.

If observations in multiple photometric bands are available, strong
constraints on the complexities of GCs can be made.
We present deep multiband optical photometry of IC$\,$4499 in the $U$, $B$, $V$, 
$R$, $I$, and $DDO$\,$51$ bands, and take advantage of this 
wavelength range for a detailed analysis of potential multiple components in this GC.

IC$\,$4499 is a little-studied  cluster lying in the intermediate-outer halo, with basic 
parameters \citep{har96, har10},  of  $\rm R_{GC} = 15.7$ kpc, $\rm M_V = -7.33$, $\rm [Fe/H] = -1.53$.  
It is a low-density cluster that allows ground-based studies to penetrate to the cluster 
center to below the main sequence (MS) turnoff, and is principally distinguished by its 
apparent relatively young age \citep{fer95} and the large number of RR Lyrae variables it 
contains \citep{fou74,cou75,cle79}; it has the highest specific frequency (numbers normalized 
to mass) of RR Lyrae variables contained in any galactic GC listed by \citet{har10} with the exception of 
the low mass cluster Palomar 13. The variables have an Oosterhoff Type I (Oo I) period 
distribution \citep{oos39}, and a substantial fraction are double-mode (RRd) oscillators, 
first studied by \citet{cle86}.  A comprehensive CCD study of the variable star population 
was made by \citet{wal96}.

From low-resolution spectroscopy of four red giant branch (RGB) stars, Russell Cannon, 
quoted in \citet{sar93}, derived a mean metallicity $\rm [Fe/H] = -1.65 \pm 0.10$, and this value 
was used in the first color magnitude diagram (CMD) study by \citet{sar93}.  A deeper 
CMD by \citet{fer95} found a best fit to the observations with an isochrone with age 11 Gyr, 
some 3 Gyr younger than the bulk of Galactic GCs on the age scale then employed.

There are two more recent studies of IC 4499. \citet{sto04} derived the cluster distance from 
single-epoch K-band observations and the $\rm M_K-log P-[Fe/H]$ relation of \citet{bon03}.  
Assuming $\rm [Fe/H] = -1.65$, he found a distance 
$(m-M)_0 = 16.47 \pm 0.04$ (statistical)$\pm 0.06$ (systematic).   
Finally, \citet{hankey10} measured a precise metallicity of $\rm [Fe/H] = -1.52 \pm 0.12$  
for 43 radial-velocity selected IC 4499 members using AAOmega on the AAT.

Using the $B$,$V$ and $I$ passbands, new CMDs for IC$\,$4499 
are presented and interpreted.  More than a decade has passed since the last CMD 
investigations of IC$\,$4499, and in that time the age of the Universe has 
been refined by the WMAP experiment (Bennett et~al. 2010); for a standard 
$\Lambda$CDM model the best-fit age is 13.7 Gyr.  Formation of galaxies and the 
first Galactic GCs is expected to have happened relatively quickly, within approximately 
the first Gyr, by $12.8 \pm 0.4$ according to \citet{mar09}.  
Modern isochrones (e.g. the BaSTI 
archive, available on-line at {\tt http://www.oa-teramo.inaf.it/BASTI/}) are used here
to investigate whether IC$\,$4499 is a ``young" GC with an age 2--4 Gyr younger
than clusters of similar metallicity \citep{fer95}, and to provide an updated
age determination of this cluster.

As the observations reach farther than the tidal radius of IC$\,$4499, an 
investigation of a possible association of IC$\,$4499
with extratidal stars is carried out.  On the basis of its supposedly young age and 
position in the Galaxy, IC$\,$4499 has been 
proposed to be associated with the Monoceros Ring, a structure whose interpretation 
is contentious \citep{hammersley10} but most likely appears to be a 
tidal stream remnant of a galaxy that merged with our own \citep{gri06,con08,cas08,che10}.  
Recently, the radial velocity of IC$\,$4499 was
shown to be within the range of the halo GCs and not inconsistent with being associated
with membership in the Monoceros Ring \citep{hankey10}.  

This paper is organized as follows:  \S2 is a description of the observational material,
\S3 is the presentation of the CMDs from which we estimate the structural parameters 
of the cluster, and compare them with theoretical isochrones from which the age, 
distance, and reddening are derived, \S4 is a discussion of the 
radial stellar distribution of IC$\,$4499, \S5 describes a search
for the existence of multiple stellar populations, and \S6 is a summary of our conclusions.

\section[]{Observations and Photometry}


\begin{table*}
\begin{scriptsize}
\centering
\caption{IC$\,$4499 Observations}
\label{phot}
\begin{tabular}{llll} \hline
Dates & Telescope & Detector & Images \\ \hline

1987 Feb 03-06 & CTIO 0.9m CFIM      & TI 800      &  6 B, 6 V, 5 R                        \\ 
1987 Mar 06-11 & CTIO 0.9m CFIM      & RCA 512x320 & 31 B, 32 V, 32 R                      \\ 
1987 Mar 29    & CTIO 1.5m CFIM      & TI 800      &  3 B, 3 V, 3 R, 2 I                   \\ 
1987 Apr 17    & CTIO 0.9m CFIM      & TI 800      &  1 B, 1 V                             \\ 
1987 Jun 05-11 & CTIO 0.9m CFIM      & TI 800      & 44 B, 44 V, 44 R                      \\ 
1988 Mar 23    & CTIO 4.0m CFIM      & TI 800      &  2 U, 2 B, 2 V                        \\ 
1988 Jul 10-11 & CTIO 0.9m CFIM      & TI 800      &  3 B, 3 V                             \\ 
1989 Apr 18    & CTIO 0.9m CFIM      & TI 800      &  7 B, 6 V                             \\ 
1990 Mar 22-24 & CTIO 1.5m CFIM      & TI 800      &  8 U, 47 B, 49 V                      \\ 
1990 Mar 24    & CTIO 4.0m PFIM      & TI 800      &  9 U, 11 B, 9 V                       \\ 
1990 Jun 20    & CTIO 4.0m PFIM      & TI 800      &  6 B, 2 V                             \\ 
1990 Jun 24-25 & CTIO 1.5m CFIM      & Tek 512     &  8 U, 36 B, 36 V                      \\ 
1992 Feb 18    & CTIO 4.0m PFIM      & Tek 2K      & 15 B, 19 V                            \\ 
1992 Jul 09    & ESO NTT EMMI        & Thomson 1K  &  7 B, 9 V, 6 R, 6 I                   \\ 
1992 Sep 19    & ESO NTT EMMI        & Thomson 1K  & 15 B, 15 V                            \\ 
1993 Mar 03    & ESO NTT EMMI        & Thomson 1K  &  9 B, 9 V                             \\ 
1993 Apr 29-30 & ESO NTT EMMI        & Tek 1K      & 18 B, 20 V, 2 R, 2I                   \\ 
1993 Jun 27-28 & ESO NTT EMMI        & Tek 1K      & 14 B, 15 V, 1 R, 1I                   \\ 
1994 Mar 07-11 & CTIO 0.9m CFIM      & Tek 2K      & 32 B, 32 V, 32 I                      \\ 
1994 Apr 26    & CTIO 0.9m CFIM      & Tek 2K      &  2 B, 2 V, 2 I                        \\ 
1994 Jun 27-28 & CTIO 0.9m CFIM      & Tek 2K      & 30 V                                  \\ 
1996 Mar 10    & CTIO 4.0m PFIM      & Tek 2K      & 20 V, 30 I                            \\ 
2008 May 03    & CTIO 4.0m Mosaic    & SITe 2Kx4K  &  4 U, 6 B, 6 V, 4 I, 4 DDO51 ($\times$8) \\ 
2009 Apr 03    & Magellan 6.5m IMACS & SITE 2Kx4K  & 22 V, 16 I ($\times$8)                \\ 
\hline
\end{tabular}
 \end{scriptsize}
 \end{table*}


\subsection{Observations}
The CCD imaging observations used here include all the observations of
IC$\,$4499 obtained by \citet[hereafter WN96]{wal96} with various telescopes at
CTIO over the period 1987--1996, and all publicly available IC$\,$4499 data that
we were able to locate; these images are now contained within a private archive
maintained by PBS.  Further observations were carried out specifically for this
project in May 2008 using the Mosaic II Imager on the CTIO Blanco 4m telescope,
and in April 2009 with IMACS on the 6.5m Magellan Baade.  The data thus span a
range from 1987 to 2009; in total, there are 1365 individual CCD images from 24
observing runs.  A summary of these data is presented in Table~\ref{phot}.  

\subsection{Photometry}
All observations were reduced using the DAOPHOT IV and ALLFRAME suite of
programs \citep{ste87,ste94}.  The photometry was calibrated to a  standard
photometric system, closely approximating \citet{lan92}, using algorithms called
CCDSTD, CCDAVE and NEWTRIAL.  These are the modules that were used to produce
the photometry discussed in \citet{ste00, ste05} and many other papers devoted
to particular star clusters.  Indeed, CCDSTD and CCDAVE have been used in
essentially their current form for more than twenty years.  NEWTRIAL is
fundamentally the same program as was used in \citet{ste96} and most of the
papers from the Hubble Space Telescope Key Project on the Extragalactic Distance
Scale \citep[see][and references therein]{freedman01}.  CCDSTD uses
synthetic-aperture photometry of established photometric standard stars to
derive transformation equations relating the observed instrumental magnitudes to
the standard photometric system.  CCDAVE employs these equations to calibrate
the instrumental measurements of those standards and of selected, isolated stars
that are intended to become local standards in the various target fields. 
Finally, NEWTRIAL uses the local standard stars and the transformation equations
to calibrate the entire corpus of observations for all stars in a target field
to the standard photometric system.

With these algorithms, it is possible to calibrate data obtained under either
photometric or non-photometric ($\rm i.e.$, thin or scattered clouds) observing
conditions.  For a night when conditions were photometric, a full transformation
solution is obtained for each filter with a form like:

$$v_{\hbox{\footnotesize observed}} = V_{\hbox{\footnotesize standard}} +
\alpha_0 + \alpha_1 X + \alpha_2 (\bmv) + \alpha_3 (\bmv)^2$$
\noindent 

\noindent where $v$ is the aperture-corrected instrumental magnitude resulting
from the CCD measurement normalized for the integration time, $X$ is the
airmass, and $V$ and \bmv\ are presumed known quantities from the literature. 
The $\alpha$'s in the transformation equation are unknown quantities to be
determined from the observations of photometric standard stars made during the
night, and are obtained by application of a robust least-squares technique.  The
instrumental magnitude $v$ has a measuring error associated with it, resulting
from Poisson photon noise, readout noise, PSF-fitting uncertainties (if
relevant) and uncertainties in the aperture correction.

In the corresponding transformation for $\bb$, the color-extinction term,
$\alpha_4 (\bmv) X$ with $\alpha_4$ usually equal to -0.016 is used.  The $\rm
U$-band transformations incorporate terms $\alpha_2 (\umb)$, $\alpha_3 (\bmv)$
and $\alpha_4 (\bmv)^2$ to deal with that awkward regime among the B-A-F stars
where a single value of \umb\ may be found in stars with three different values
of \bmv.  Spatial terms $\alpha_5 x$ and $\alpha_6 y$ involving the $(x,y)$
coordinates of the star within the digital CCD image are used for large-format
CCDs and for mosaic cameras.  

Observations made on non-photometric occasions can be used to reduce the random
photometric errors among stars in a single target field, producing tighter
sequences in color-magnitude and color-color diagrams.  For images taken on
non-photometric nights, no extinction correction or universal zero-point is
derived.  Instead, we use a transformation equation of this general form:
$$v_{\hbox{\footnotesize observed}} = V_{\hbox{\footnotesize standard}} +
\zeta_i + \alpha_2 (\bmv) + \alpha_3 (\bmv)^2 + \ldots.$$
\noindent A unique zero-point, $\zeta_i$, for each CCD image $i$ is derived from
standard stars contained within that image.  The color-correction terms can be
derived from CCD images containing multiple standard stars with a range of color
and/or from observations made with the same equipment during
photometric nights of the same observing run,

The program CCDAVE inverts the calibration equations: the instrumental
magnitudes (e.g., $v$) are still regarded as observed quantities with associated
observational standard errors;  the $\alpha$'s are now regarded as known
constants; and the star's calibrated magnitude on the standard system $\vv$
(etc.) is the unknown to be determined by a robust least-squares statistical
adjustment.  Statistical estimates of the single optimum values of the
magnitudes in other bandpasses, such as $U$, $\bb$, $R$, and $I$ are
obtained in exactly the same way at the same time.  Since standard-system colors
must be known for the color terms of the various photometric transformation
equations, determination of these standard-system magnitudes must be iterative
in nature; a neutral color is initially assumed for the star and substituted
into the various transformation equations associated with the different
observations of it, and as estimates of the star's standard-system magnitudes
improve, the color indices employed in the transformation equations are refined
and the magnitudes are redetermined.  The coefficients of the color terms are
small---a few hundredths of a magnitude per magnitude for a filter that is a
reasonable approximation to the standard bandpass---and the convergence to a
satisfactory solution is almost never a problem.

CCDAVE can derive standard-system magnitudes for the standard stars exactly as
it does for the target stars of unknown photometric properties.  The inputs to
the program are only the observed instrumental magnitudes and the transformation
equations.  CCDAVE has no prior knowledge of the standard photometric indices
or, indeed, which stars are standards and which are targets.  A comparison
between the output results for the standard stars produced by CCDAVE and the
published values for the same stars that were used as input to CCDSTD,
therefore, provides a clean test of the end-to-end validity of the methodology. 
As of the date of the submission of the revised version of this article, a total
of 2,076 datsets have been homogeneously analyzed with CCDSTD and CCDAVE; among
these are 263 that include observations of IC$\,$4499.  


\begin{figure}
\includegraphics[width=9cm]{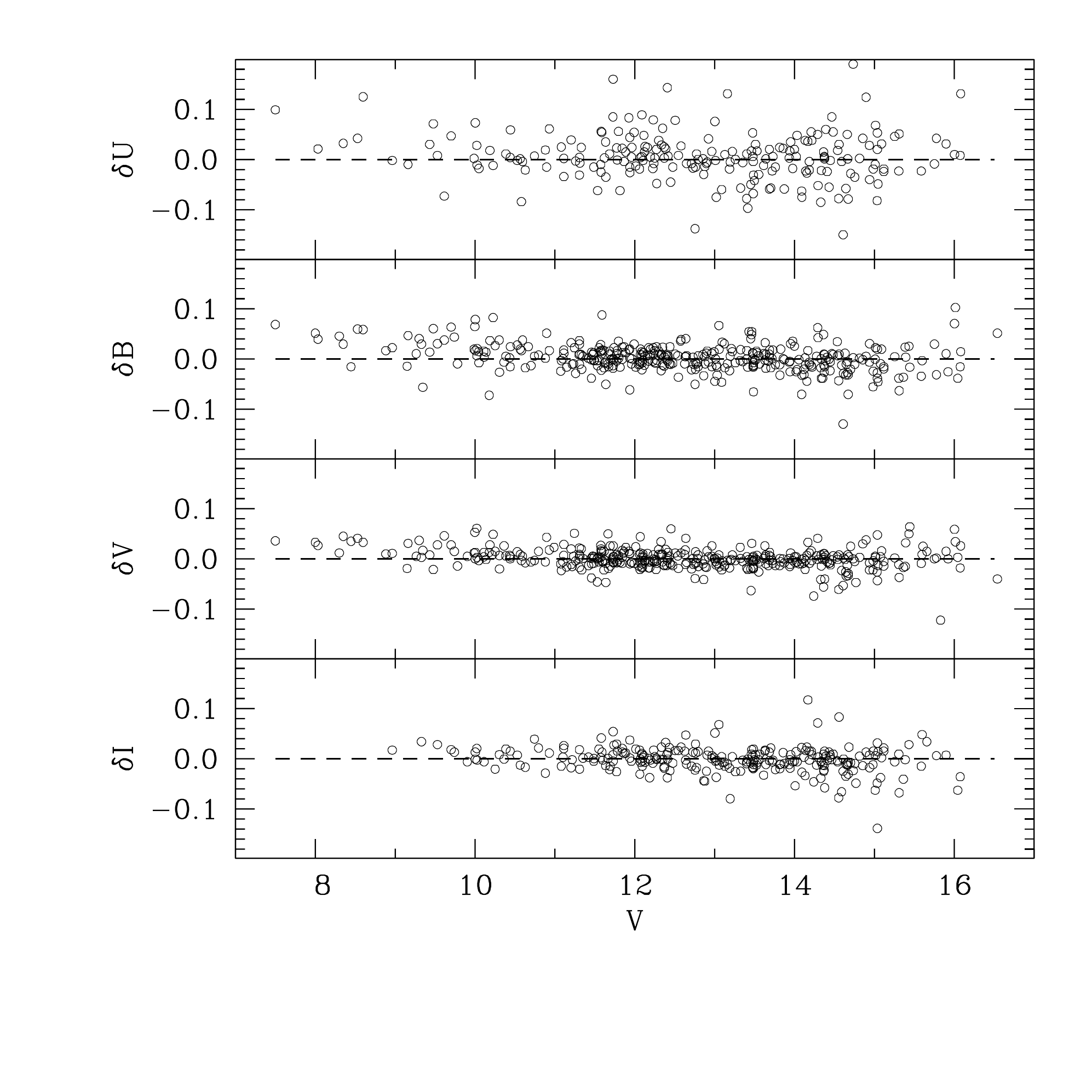}
\caption{The difference in $U$, $\bb$, $\vv$, and $I$ for stars
in standard-system magnitudes and the stars obtained here, plotted against 
apparent visual magnitude.  From 215 to 376 stars are plotted, depending
upon the filter.
}
\label{do1}
\end{figure}


\begin{figure}
\includegraphics[width=9cm]{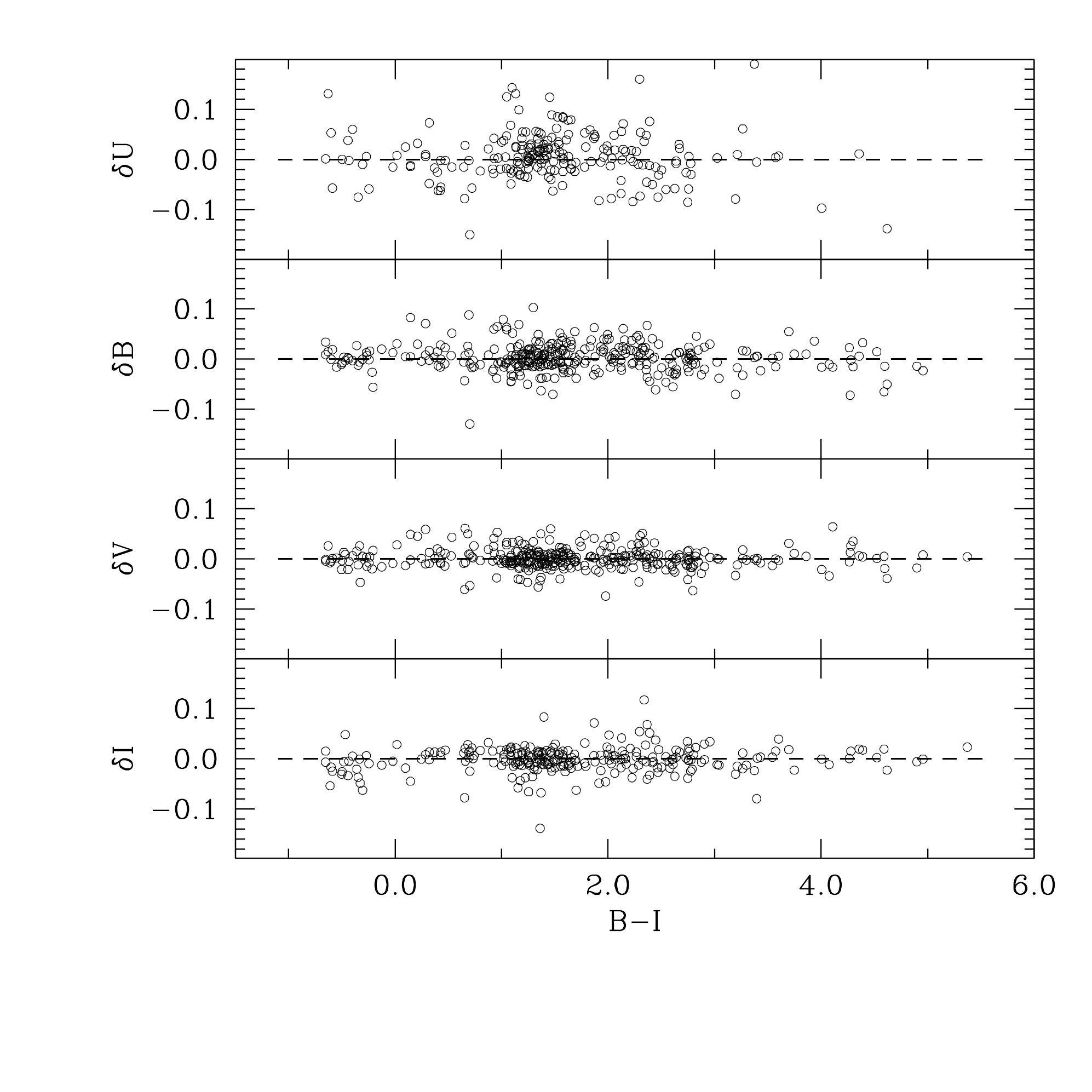}
\caption{The difference in $U$, $\bb$, $\vv$, and $I$ for stars
in standard-system magnitudes and the stars obtained here, plotted 
against $\rm(B-I)$ color. }
\label{do2}
\end{figure}


Figure~\ref{do1} and Figure~\ref{do2} show the magnitude differences between the
standard-system magnitudes produced by CCDAVE and the published values for the
fundamental standard stars---those of \citet{lan73, lan83, lan92} and
\citet{graham81, graham82}\footnote{Apart from modest additive zero-point
adjustments, $-0.002\pm0.002$, $-0.002\pm0.001$, and $-0.002\pm0.001$ in $\rm
U$, $\bb$, and $\vv$, respectively, to place the \citet{lan73} results on the
same system as \citet{lan83, lan92}, based upon 119 stars in common, we find no
evidence within our data to suggest that these studies do not define the same
photometric system, within the errors.}.  Figure~\ref{do1} shows the magnitude
residuals (in the sense published minus ours) plotted versus apparent visual
magnitude, while Figure~\ref{do2} plots the same residuals against \bmi\ color,
serving as a proxy for effective temperature or, equivalently, spectral class.  
The stars shown here have been observed on a minimum of five photometric
occasions, and have a standard error of the mean magnitude smaller than
0.02~mag.   The weighted mean differences between the published results and ours
are $+0.0035\pm 0.0024$, $+0.0028\pm0.0011$, $-0.0009\pm0.0008$ and
$-0.0009\pm0.0012$~mag in $U$, $\bb$, $\vv$, and $I$, respectively,
based upon 215, 361, 376, and 253 stars.  (For completeness, we report here a
similar difference of $-0.0002\pm 0.0009$~mag, based on 244 stars, in the $\rm
R$ filter; the $R$ band plays no role in our study of IC$\,$4499.) Expressed
as a standard deviation, the average standard residual of one star is 0.032,
0.020, 0.014, and 0.017~mag in $U$, $\bb$, $\vv$, and $I$.  Together,
these two figures demonstrate that our version of the standard photometric
system matches the definitive published results for the same stars with a high
degree of fidelity over a dynamic range of $\sim$600 in flux ($9 \ltsim V \ltsim
16$) and over the full available range of temperature. 


Considering now the observations of IC$\,$4499, CCDSTD and CCDAVE are 
employed to define a sequence of local
photometric standards in each target field of scientific interest, by transfer
of photometric calibration from fundamental photometric standards in other parts
of the sky.  NEWTRIAL transfers the calibration
to the complete set of instrumental magnitudes measured by the DAOPHOT/ALLFRAME
suite of algorithms for all stars in a particular target field---in the present
case, IC$\,$4499.  NEWTRIAL considers results from all observing nights and runs
simultaneously.  It assembles the totality of observed instrumental magnitudes
for any given star, associates each measured instrumental magnitude with the
transformation equation appropriate to the night, detector, filter, and image
from which it was obtained, and uses a robust least-squares adjustment to
estimate the single best mean magnitude in each photometric bandpass that
optimally explains the corpus of observations in the corresponding filter.

Some exposures using the $DDO$\,$51$ filter were also taken.  Observations
in this filter were standardized by using the approximation 
\\ \\ 
$DDO$\,$51 = 0.67*V + 0.33*B$
\\ \\ 
which was then used to define ``standard" $DDO$\,$51$ magnitudes for the
local standard stars.  Observed indices were fitted to these values in the usual
way. Then we defined
\\ \\ 
$\Delta = DDO$\,$51 - 0.67*V - 0.33*B$,
\\ \\ 
which is a good gravity-sensitive index for red stars, as clearly shown by \citep{teig08}.  
In particular, $DDO$\,$51$ can be useful in separating globular cluster Red Giant 
Branch (RGB) stars from foreground dwarfs.  

Additionally we note that IC$\,$4499 has a high southern declination and 
as a consequence its airmass as
viewed from telescopes at the mid-latitude Chilean observatories is always
rather high and nearly constant,  $\sim$1.8--2.0.  The cluster observations
themselves cannot be used to self-calibrate extinction during a night's
observations, as its own extinction hardly changes. 

We find significant offsets between the standardization performed here
and reductions of the data presented by
WN96.  In particular, the following relations are found for
42 of 55 local standards from WN96:    
\\ \\ 
$V$ (here) = $V$ (WN96) + 0.042 
\\ 
$B$ (here) = $B$ (WN96) + 0.044 
\\ 
$I$ (here) = $I$ (WN96) + 0.017 
\\ \\
and therefore:
\\ \\ 
$(B-V)$ (here) = $(B-V)$ (WN96) + 0.002 
\\
$(V-I)$ (here) = $(V-I)$ (WN96) + 0.025 
\\
$(B-I)$ (here) = $(B-I)$ (WN)96 + 0.027 
\\ \\ 
We believe that the addition of extra observations on photometric nights and
the use of vastly greater number of (secondary) photometric standards made in
the present study has allowed a more accurate calibration.   However, and
importantly, the $B-V$ offset between the two studies is almost negligible,
and so the extensive discussion on reddening in WN96 is still relevant.  As
photometry in the $V$, $B$, and $I$ bands is more extensive than for
the other bands, only the photometry in these filters will be used when
calibration to a standard photometric system is of importance.  

For each object the detected error, CHI of fit, and roundness parameter are used
to help eliminate poorly measured stars, unresolved blends, and non-stellar
objects.  Intensity and magnitude mean magnitudes were calculated for the RR
Lyraes using codes which fitted Fourier series to the data.  The extensive
IC$\,$4499 RR Lyrae variable star observations have been combined with earlier
photographic observations for a discussion of period changes in OoI and OoII
clusters \citep{kun11}. 


\begin{table*}
\centering
\caption{Star Counts (log N) as a function of $V$ magnitude}
\label{completeness}
\begin{tabular}{lllllll}\hline
Magnitude & 21.0 - 21.5 & 21.5 - 22.0 & 22.0 - 22.5 & 22.5 - 23.0 & 23.0 - 23.5 & 23.5 - 24.0 \\
Annuli (arcsec) \\ \hline
 
0  -  50 &  2.42 &  2.26 &1.95 & 1.26 & 0.30  & 0.0 \\
50 - 100 & 2.77  & 2.79  & 2.75  & 2.42  & 1.70  & 0.78 \\
100 - 140 & 2.53  & 2.66  & 2.67  & 2.60  & 2.28  &1.48 \\
140 - 210 & 2.53  & 2.64  &2.77  & 2.84  & 2.67  & 2.08 \\
210 - 300 & 2.26  & 2.40  & 2.54  & 2.64  & 2.57  & 2.07 \\
300 - 450 & 2.07  & 2.24   & 2.28 & 2.41 & 2.40 & 1.86 \\
\hline
30 - 50 & 2.24 & 2.11 & 1.85 & 1.15 &  0 & 0 \\
50 - 70 & 2.38 & 2.36 & 2.18 &1.83 &1.80 & 0 \\ \hline
\end{tabular}
 \end{table*}



\section{C-M Diagram of IC$\,$4499}

\subsection{Completeness}
The fitting of theoretical isochrones to data requires a knowledge of the 
completeness of the star sample to avoid introducing any biases, particularly 
in the critical main sequence turnoff (MSTO) region.   The catalogue extends 
several magnitudes below the MSTO region.  As there will be no discussion 
of the main sequence luminosity function, it is not necessary
to perform a rigorous analysis at faint magnitudes.  In particular, we remind the 
reader that our body of data consists of 1365 CCD
images obtained with at least twelve different equipment setups on five different
telescopes.  The individual images were centered on many different locations
within the cluster field, and the various images were subject to many different
conditions of seeing, guiding, and sky brightness.  Any given star can have as
many as 357 observations in a single filter, or as few as one.  To map out the
detection completeness as a function of magnitude, color, and position on the
sky would therefore be a prohibitively expensive endeavor.  We thus map
out the zone of magnitude/position space where completeness issues do
not unduly complicate matters by a simpler and more approximate methodology.

The limits of sample completeness are estimated by considering the cluster
luminosity function in a series of radial zones.  These zones were chosen to
cover the cluster extent, with roughly similar numbers of stars in each zone.
It is presumed that the worst incompleteness will occur in the innermost zone, where
the crowding is most severe, and possibly also in the outermost zone, where the
total available exposure time is least.  

Our procedure is to plot the logarithm
of the star counts against apparent magnitude.  The luminosity functions of the
outermost two zones are compared: over the magnitude range where they are
closely parallel, it is presumed that both are comparably complete.  Near the
faint end, the apparent magnitude at which they begin to diverge is where one or
the other is beginning to be incomplete.  That done, the next zone---counting
inward---is compared to the sum of the two outermost zones, and the
magnitude extent over which the luminosity functions are closely parallel is
again determined.  Then the next radial zone is compared to the sum of the
three outer zones, and so on.  In this way, the apparent magnitude of
incipient incompleteness as a function of radial distance from the cluster center
is mapped.  With this approach, as one moves inward, the magnitude range over which
the luminosity functions may be compared becomes increasingly restricted as the
onset of incompleteness moves brightward.  At the same time, however, the
star-counting statistics are improved, as each individual radial zone is compared
to the sum of all zones lying farther from the cluster center.

\begin{figure}
\includegraphics[width=9cm]{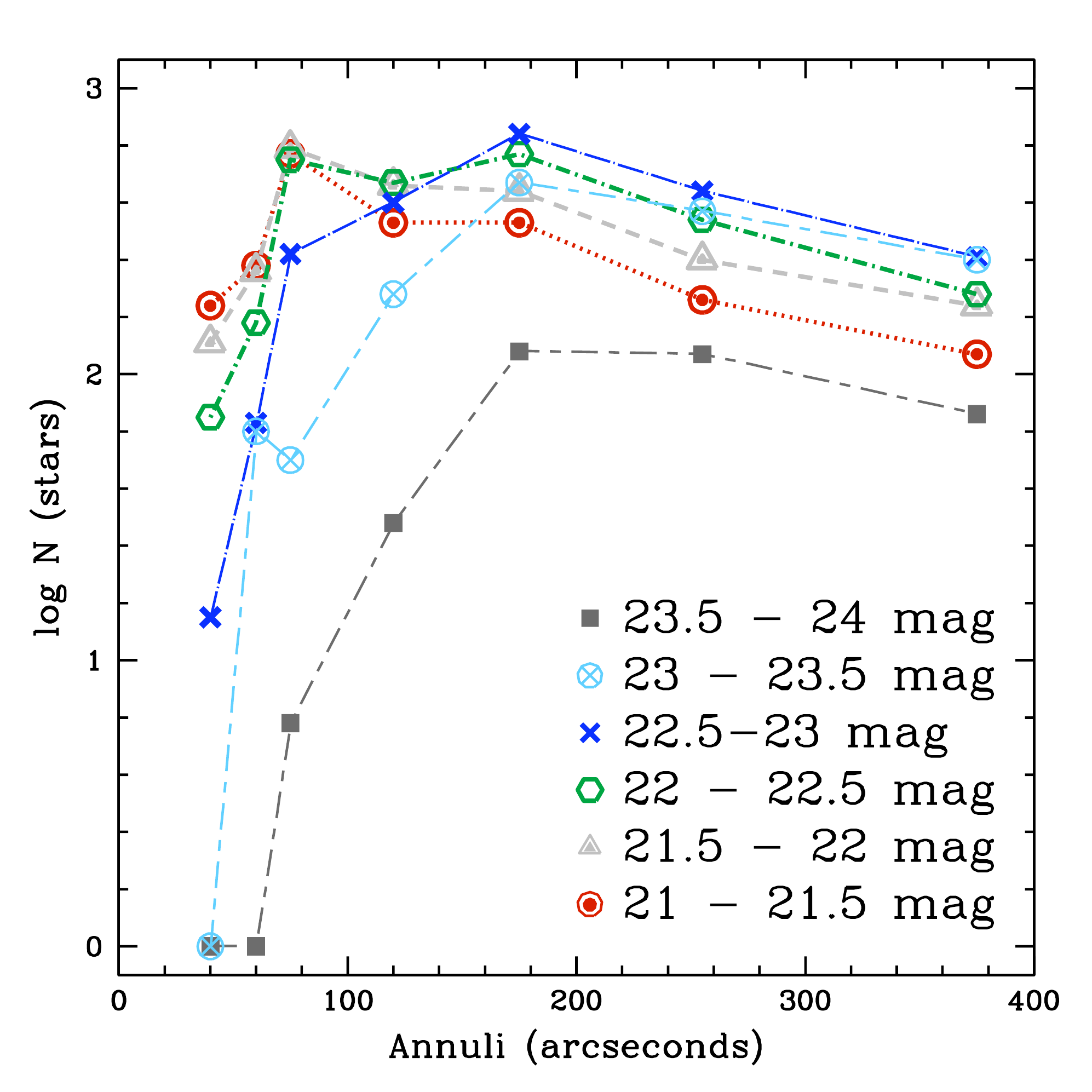}
\caption{{\it Left:}  The number of stars per radial zone for IC$\,$4499.  The
different symbols represent different magnitudes bins, which are specified
in the bottom right corner.}
\label{LF}
\end{figure}

Here 0.5 mag wide bins are chosen and a color range sufficient to contain the 
MS stars.   The \vv,\vmi \  CMD is used, as this is deeper and tighter than the others, 
primarily due to the high-resolution Magellan data.  Zones with differing radii 
are chosen (see Table~\ref{completeness}), and the luminosity functions for the different
zones are then slid until they overlap, starting with the outer zone then 
sliding and adding the inner zones one by one.  These results are illustrated in 
Figure~\ref{LF} and Table~\ref{completeness} where the columns contain (1) the radial zone and 
(2)--(5) the number of stars in each magnitude bin.

Because the 
completeness as a function of magnitude changes rapidly as 
one approaches the cluster center, we also experimented with zones of 30-50 
arcsec and 50-70 arcsec.  For the six zones constructed here the field star contamination is 
very small, and can be neglected.  Furthermore, the completeness statistic for 
an ALLFRAME-produced catalogue remains close to 100\% as the catalogue limit 
is approached and then rapidly falls to zero at fainter magnitudes, in contrast to a 
single reduction of stacked frames (cf., for instance, Stetson 1991).    From
Table~\ref{completeness}, the $\sim$90\% completeness level is derived and shown
in Table~\ref{comp90}.  

In conclusion, for our study of the populations of resolved stars brighter than 
the MSTO ($V \sim 20.5$) as a function of radius, outside a radius of 
50 arcsec completeness corrections are negligible (Table~\ref{comp90}).  
Hence, fitting isochrones 
to a similar sample should be safe.  In principle, when fitting cluster observations with
isochrones, it is not necessary for the sample to be complete, but that the sample 
is statistically significant.  This is because the isochrones are fit to the shape of the 
CMD.  The completeness is more
important for star counts, LFs,  and for the luminosity profile.

\subsection{Reddening}
IC$\,$4499 is an outer halo cluster, one of 35 such according to the 
$\rm R_{GC}$ $>$ 15  kpc definition of \citet{van04}.  With 
$(l,b)$ = (307\hbox{$.\!\!^\circ$}35, $-$20\hbox{$.\!\!^\circ$}47) 
we view the cluster on the far side of our Galaxy and through a considerable 
amount of dust.  WN96 calculated the reddening for IC$\,$4499 using 
four methods, two of these are dependent on the cluster metallicity for which 
they chose ${\rm [Fe/H]} = -1.65$, and the other two involved colors 
of the RR Lyrae variables and the colors of the instability-strip boundaries.   
All methods were consistent, with the result $E(B-V) = 0.22 \pm 0.02$.   
To this we add the result from use of the \citet{sch98} reddening maps which
show that a 5 arcmin radius centered on IC$\,$4499 has 
$E(B-V) = 0.224 \pm 0.004$ (statistical).  Regions centered approximately 
15--20 arcmin N and S have slightly higher reddening, $E(B-V) \sim 0.24$ 
and 0.26 respectively, but elsewhere within 15 arcmin of the center of 
IC$\,$4499 the reddening is within the range $E(B-V)$ $=$ 0.21--0.23.   
Given the relatively small angular size of the cluster, there appears to be no need to 
apply variable reddening corrections, and indeed we do not find any evidence for 
their necessity.  However small scale variations in reddening at the few percent 
level ($\sim$0.01 mag) cannot be ruled out.     In summary, we find 
from the several methods available that $E(B-V) =  0.220 \pm 0.005$, where 
the error is derived from the scatter of the individual estimates. As systematic 
errors in the various calibrations likely dominate, we will adopt $E(B-V)$ = 0.22 $\pm$ 0.02. 


\begin{table}
\centering
\caption{$V$-band Magnitude Completeness Limits }
\label{comp90}
\begin{tabular}{ll} \hline
Annuli (arcsec) &  90\% Limit \\ \hline

0-30   &  $<$ 21.25\\
30-50   & 21.25 \\
50-70   &  22.0 \\
70-100   & 22.25 \\
100-140   & 22.75 \\
140-450   & 23.0 \\ \hline

\end{tabular}
 \end{table}


In order to compare the observed CMD to isochrones, and to fit the CMD in colors 
other than $(B-V)$, we solve equations 1 and 3 of \citet{car89} for a variety of 
$\rm R_V = A_V/E(B-V)$.  For example, with $R_V = 3.1$ and $E(B-V)=0.22$, 
$E(V-I)=0.30$ whereas for $R_V=3.3$ and $E(B-V)=0.22$ then $E(V-I)=0.31$ (here we
assume that the $I$ photometric band corresponds to an effective wavelength of 800$\,$nm;
cf.\ Kron, White \& Gascoigne 1953; Cousins 1976; Landolt 1983).
Within reasonable limits, $R_V$ is an adjustable parameter. However in fitting the 
isochrones (see below) we found no reason not to choose $R_V = 3.1$.

\subsection{CMDs of IC$\,$4499}

\begin{figure*}
\includegraphics[height=13cm]{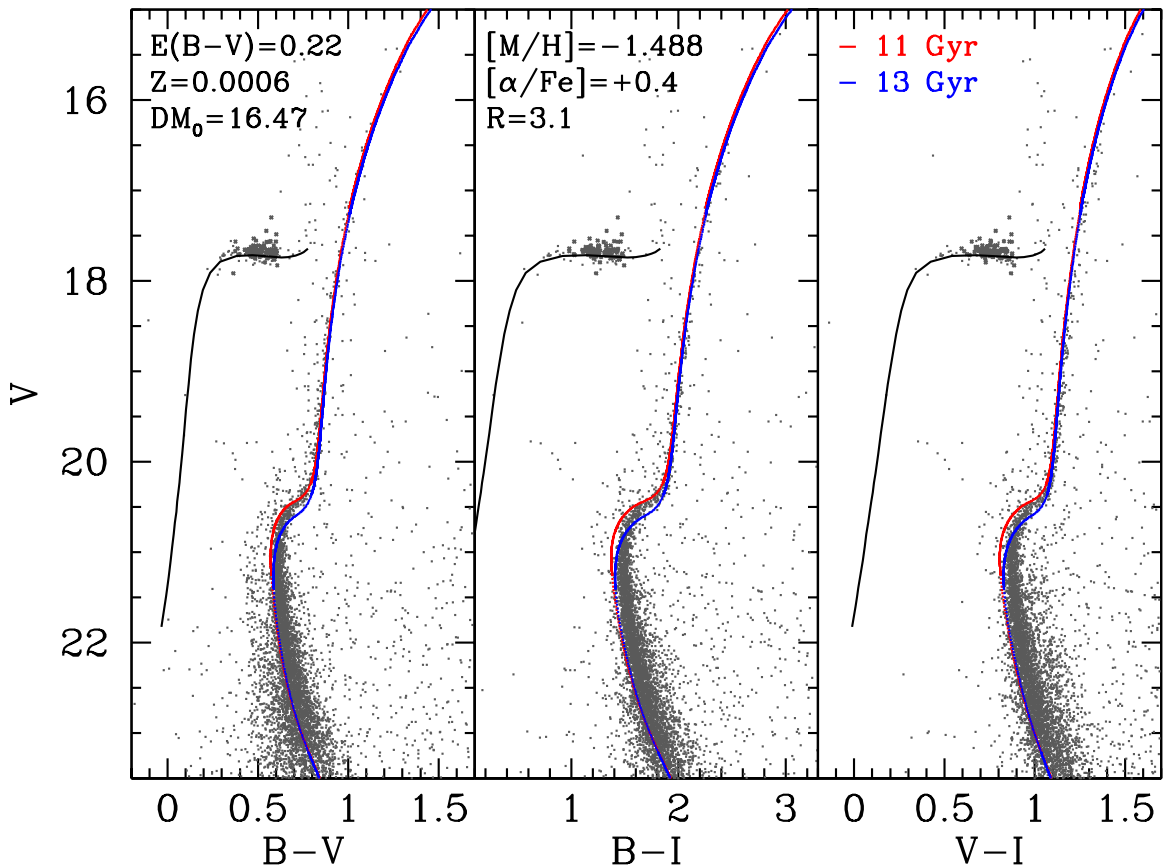}
 \caption{IC$\,$4499 Color Magnitude Diagrams to the 
 \vv,\bmv \ CMD (left), \vv,\bmi \ CMD (center) and \vv,\vmi \ CMD 
 (right) are shown.  Over-plotted are isochrones from the BaSTI 
 database and correspond to canonical, $\alpha$-enhanced calculations 
 for ages of 11 and 13 Gyr, with other parameters as displayed on the plots.
 The RR Lyrae variables are designated by crosses.}
\label{cmd}
\end{figure*}

The CMD of IC$\,$4499 is shown in Figure~\ref{cmd}, with three different color baselines.   
The morphology has been described in previous investigations, with the exception that here 
the MS extends to $V \sim 24$ with high precision.   The short horizontal branch is well
populated in the region of the instability strip; hence many of these stars are 
RR Lyrae variables, marked by crosses in Figure~\ref{cmd}.   
There is considerable contamination of the cluster RGB with field stars, due to the cluster 
position in the sky.   

Figure~\ref{cmd} shows the BaSTI isochrones set with the following parameters:  
canonical, $\alpha$-enhanced, Z = 0.0006, [$\alpha$/Fe] = +0.4, $\rm [M/H]$ = $-$1.49, 
$E(B-V)$ = 0.22, $R_V$ = 3.1, $\rm DM_0$ = 16.47, and these isochrones are 
simultaneously fitted on the \vv,\bmv, \vv,\bmi \ and \vv,\vmi \ CMDs.  The ZAHB locus 
with these same parameters is also shown.  Because the 
stellar models used in the present analysis do not account for the updated conductive
opacities provided by \citet{cassisi07}, a shift
of $+0.05$ mag in \vv \ to the BaSTI ZAHB is applied as discussed by the quoted
authors.  We established the distance modulus by matching 
the isochrones to the horizontal branch stellar distribution, and find that it is identical 
to that found by \citet{sto04}.  The metallicity is consistent with spectroscopy of 
RGB stars from \citet{hankey10} from which the cluster metallicity was 
found to be \feh = $-1.52 \pm 0.12$.  The $\alpha$-element enhancement is assumed 
to be similar ([$\alpha$/Fe] = +0.4) to that for halo globular clusters in general.

The data shown in Fig.~\ref{cmd} reveals that theoretical isochrones with 
age 11 Gyr and 13 Gyr appear to properly bracket the cluster photometry. Therefore, 
we estimate an 
age of 12 $\pm$ 1 Gyr  for IC$\,$4499.  It is worth noting that present result does not 
support the younger age of IC$\,$4499 
found by \citet{fer95} compared to the majority of Galactic GCs.  On the contrary, our age is 
coeval with them \citep{mar09}.  

The difference in the age estimate 
with respect to that of \citet{fer95}, can be - at least partially - explained as a consequence of 
the use of a more updated and reliable theoretical framework. 
However, the use of a different, updated set of isochrones such as those provided
by \citet{vand00} and \citet{dott07}, does not affect our final cluster age determination. In fact,
all state-of-the-art isochrones such as the BaSTI set give ages that are consistent with the 
WMAP age of the Universe of 13.7 Gyr \citep{ben10}, and provide a quite similar age ranking 
for the Galactic GCs system \cite[see, e.g.][and references therein]{deangeli05, mar09,cassisi11}.

We emphasize that care must be taken when comparing relative ages with those in the 
older literature, as the range of ages for the majority group of older GCs with no age-metallicity 
relation is under 1 Gyr \citep{mar09} compared to the several Gyr commonly proposed in the 
past, where uncertainties in the distance scale and magnitude-metallicity relation for RR 
Lyrae variables propagated into the ages, as discussed for instance by \citet{buo89}
and \citet{wal92}.

\section{Tidal radius and extra-tidal stars}
There is increasing evidence that some
globular clusters now part of our Galactic GC system formed in dwarf 
galaxies that subsequently---over a Hubble time---merged with our own Galaxy, with 
the present-day merger example of the Sagittarius dwarf and its central GC M54 
and other associated clusters being a case in point \citep{bel08,car10a}.  
Several GCs have been associated with stellar streams in the Galactic halo 
that are presumably remnants of earlier mergers, e.g. NGC 5053 \citep{lau06},
NGC$\,$5466 \citep{grij06} and Terzan$\,$5 \citep{fer09,origlia10}, 
while others are surrounded by extra-tidal stars 
e.g., NGC$\,$1851 and NGC$\,$1904 \citep{ols09,carb10} or distinct tidal tails such 
as that for Palomar~5 \citep{ode10, roc02} indicating ongoing dynamical 
evolution.  In a spectroscopic survey of a limited number of giant candidates
in the IC$\,$4499 field, Hankey \& Cole (2010; also Hankey 2011, private communication)
found several stars that appear to share the cluster's radial velocity
despite lying more than 24 arcminutes from the cluster center.

The tidal radius of IC$\,$4499 was determined to be 12.35 arcmin by \citet{tra92}.
The structural parameters for most of the clusters included in \citet{tra92}
were derived via surface-brightness measurements
from photographic plates and/or small-format CCD images, after manual removal of
areas affected by stars believed to belong to the foreground.  In the case of
IC$\,$4499 alone, the cluster profile was derived via numerical differentiation of
previously published photoelectric photometry; the authors do not mention any
correction for field-star contamination for this cluster.  Due to its location
in the sky, ($l^{II}$,$b^{II}$) = (307,$-$20), 
IC$\,$4499 is unfortunately seen through a
fairly rich Galactic foreground: simulated star counts from the Besan\c{c}on
model \citep{robin03} 
(see particularly their Fig.~6) predict approximately
3000 field stars per square degree with $\vv$ brighter than 20$\,$mag.  Most of these
stars are expected to have colors similar to the RGB and the turnoff region.

Our data for IC$\,$4499---which appear to be appreciably better than any that were
previously available---extend to a maximum distance of 33 arcmin from the
cluster center, and we have complete coverage out to a distance of 14 arcmin.
Therefore we should be able to derive an independent and possibly more
definitive measure of the cluster's tidal radius than has hitherto-fore been
possible.  

We have experimented with a new methodology for distinguishing the cluster
density profile from foreground-field contamination, which we will describe
here in some detail.  We begin by identifying an ``acceptance region'' 
or ``box'' in the color-magnitude plane where the vast majority of cluster
members are found.  Our assumption will be that the number of true cluster
members falling outside the acceptance region will be small compared to those
inside.  The number of field stars contained within the acceptance box may
be small, relative to the number falling outside it, but we do not expect that
number to be negligible.  We will, however, assume that neither the cluster
population nor the field population changes significantly with position within
our overall field of view.  The surface density of field stars falling outside
the box in the CMD will be easy to estimate quantitatively, because we expect
contamination of this sample by cluster stars to be small and---in
particular---negligible at large distances from the cluster.  We will then
examine the ratio of star counts {\it inside\/} the acceptance region to counts
{\it outside\/} the acceptance region as a function of distance from the cluster
center.  We will take the asymptotic limit of this ratio at large radial
distance as representing the number ratio of field stars inside the box to field
stars outside the box.  The surface density of field stars inside the box thus
derived can then be subtracted from the radial profile of star counts to yield
the decontaminated cluster profile.  

\begin{figure}
\includegraphics[width=9cm]{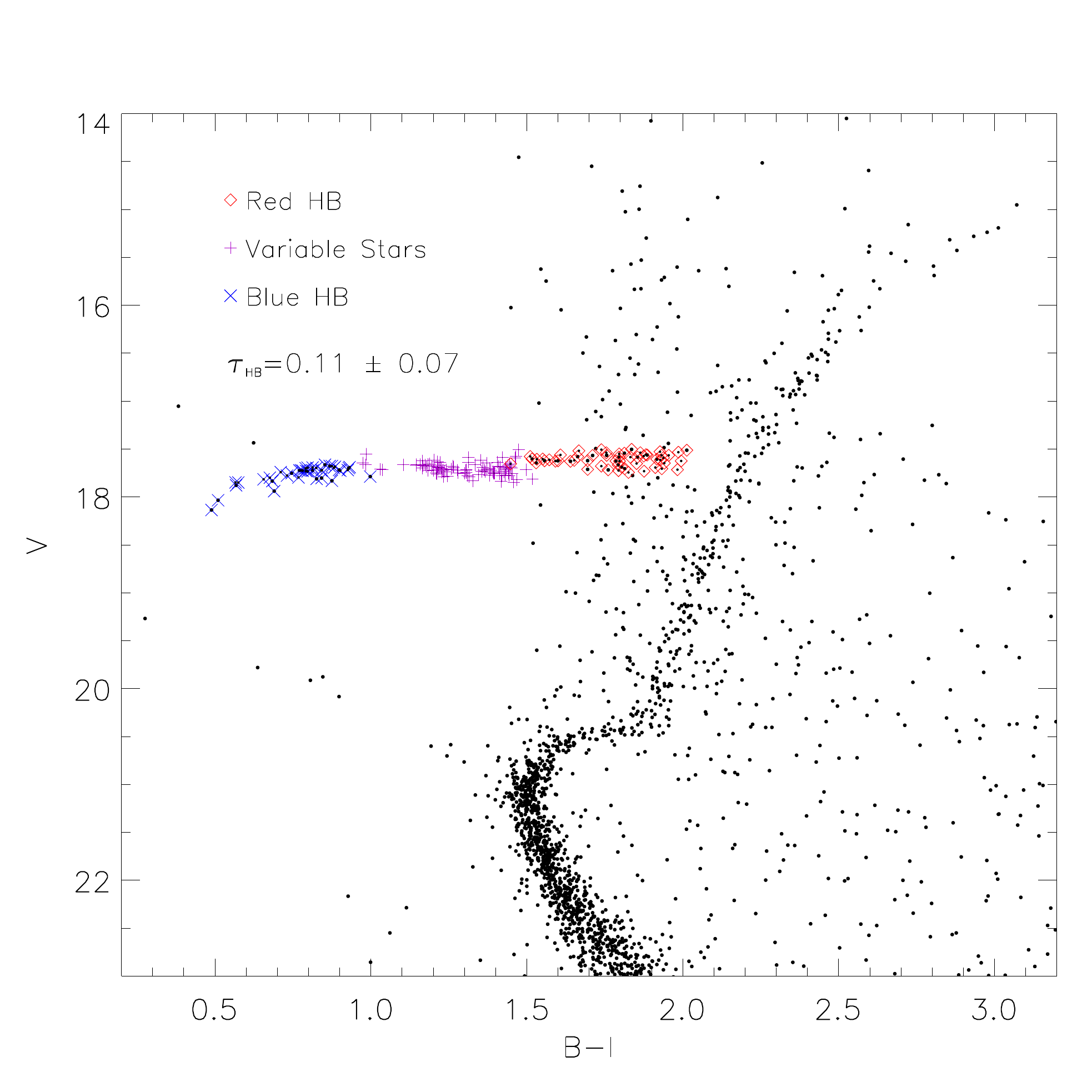}
 \caption{The IC$\,$4499 $V,B-I$ Color Magnitude Diagram in which
only stars with small
photometric errors and within 10\arcsec$\le$ r $\le$ 180\arcsec are plotted.}
\label{bmiv}
\end{figure}

\begin{figure}
\includegraphics[width=9cm]{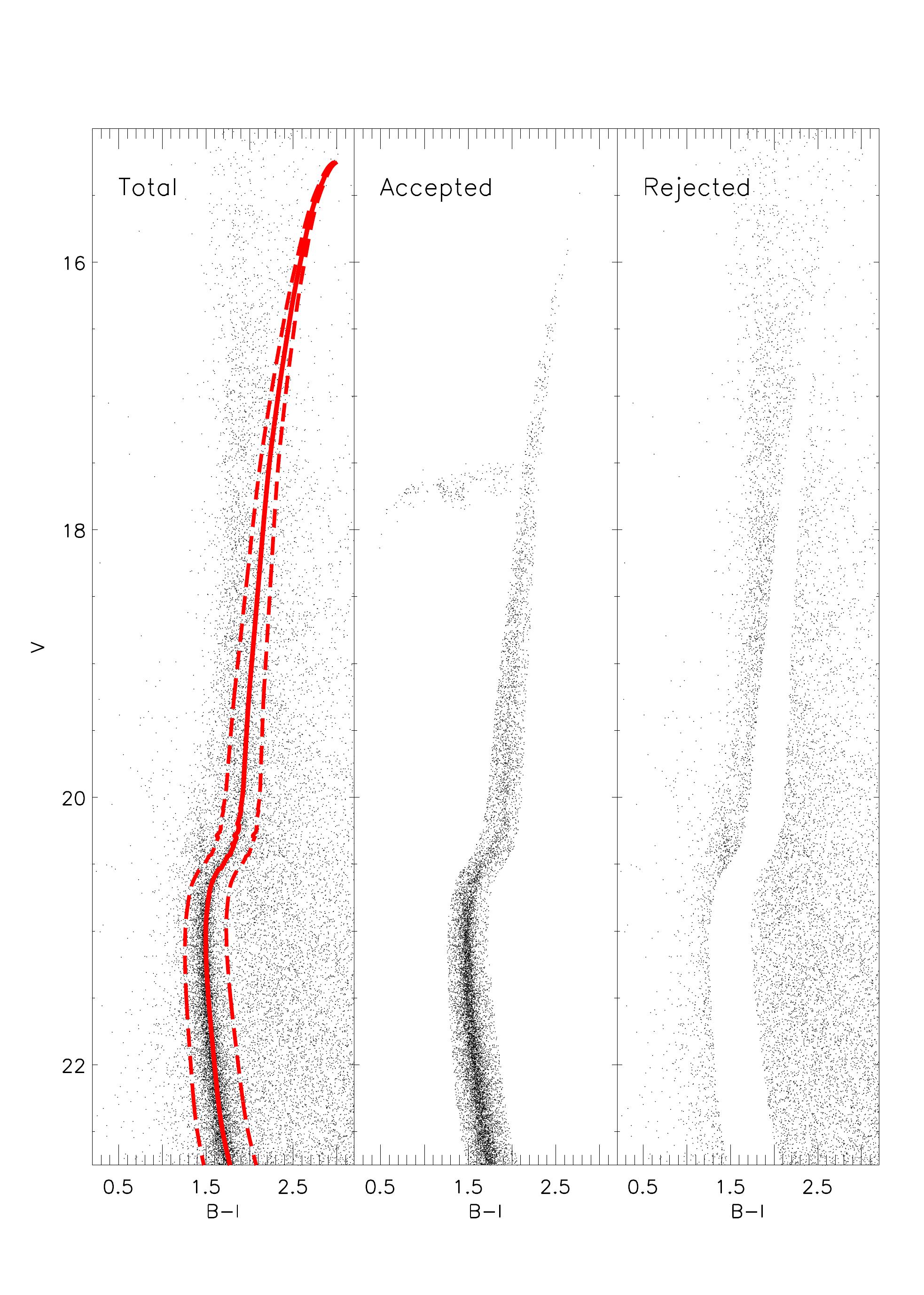}
\caption{{\it Left:}  The defined acceptance region for IC$\,$4499 cluster members is shown
with all observed stars from our stellar catalogue over-plotted.  {\it Middle:}
Only the candidate IC$\,$4499 cluster members are shown.  
{\it Right:}  The probable IC$\,$4499 field stars are shown. }
\label{ridger}
\end{figure}

To isolate that part of the color-magnitude diagram where
cluster members dominate, we plotted the stars with the smallest intrinsic
photometric errors lying within the annulus 10\arcsec$\le$ r $\le$ 180\arcsec
centered on the cluster (see Figure~\ref{bmiv}).  A cubic spline was fitted
to the cluster ridge line and an acceptance region was defined where the
majority of cluster stars are found to lie.  The left panel of
Figure~\ref{ridger} displays the \vv,\bmi \ CMD of the entire sample of stars
with no selection in radial distance and with a very mild selection in
photometric error ($\sigma_{\it BVI}$$\le$0.25 mag).  The solid line shows the
ridge line,  while the dashed ones display the acceptance region.  This
box has a width in color of 0.01 mag close to the tip of the Red Giant Branch
(RGB) and increases to 0.3 mag at magnitudes fainter than the main sequence
turn-off (MSTO).  At visual magnitudes fainter than 22.75 mag, the acceptance
region stops because the photometric errors increase and the ridge line is less
accurate.   

Using a \vv,\bmi \ CMD has the advantage of incorporating three 
independent measurements, thus limiting the possible occurrence of spurious 
detections.  Further, \bmi \ colors have 
a stronger sensitivity to the effective temperature when compared with \bmv \ and 
\vmi \ colors, and in turn give a more robust estimate of the ridge line.

The middle panel of Figure~\ref{ridger} shows the cluster HB stars as well as
the stars located inside the acceptance region.  The candidate IC$\,$4499 stars 
in the acceptance region encompass a rather generous color range, since a 
fraction of them might be affected by differential reddening.  The right panel of 
Figure~\ref{ridger} shows the CMD of candidate field stars (rejected)
with their typical peak in color around (\bmi)=1.8--2 mag. 
The group of stars located at \vv$\sim$20 and (\bmi)$\sim$1 includes cluster blue 
stragglers (BSs) and also chance optical blends of stars in the crowded regions of
the cluster.  Note that we do not count these among the cluster stars, which will
result in a (very) minor underestimate of the number of true cluster members in the
innermost one or two radial bins.  This will have no significant effect whatsoever
on our inferences for the outer part of the cluster profile.  The BSS will be included 
in a study of the IC$\,$4499 RRL and SX Phe stars by Nemec et al (in preparation).

With the acceptance region defined, we can now consider the radial density
profiles of the ``accepted'' and the ``rejected'' stars.  The center of the
cluster lies some 887 arcseconds from the eastern edge of our survey area,
1386 arcseconds from the west edge, 842 arcseconds from the north edge, and
1508 arcseconds from the south edge.  Therefore, we have a complete star sample
only out to a radius of 842 arcseconds or 14 arcminutes; from 14 to 34 arcminutes
we have a representative but incomplete sample.  Accordingly, the field is
divided into concentric annuli with an outermost limit at
$r\sim$800\farcs; this is slightly larger than the aforementioned
estimate of the tidal radius of the cluster.

\begin{figure}
\includegraphics[width=9cm]{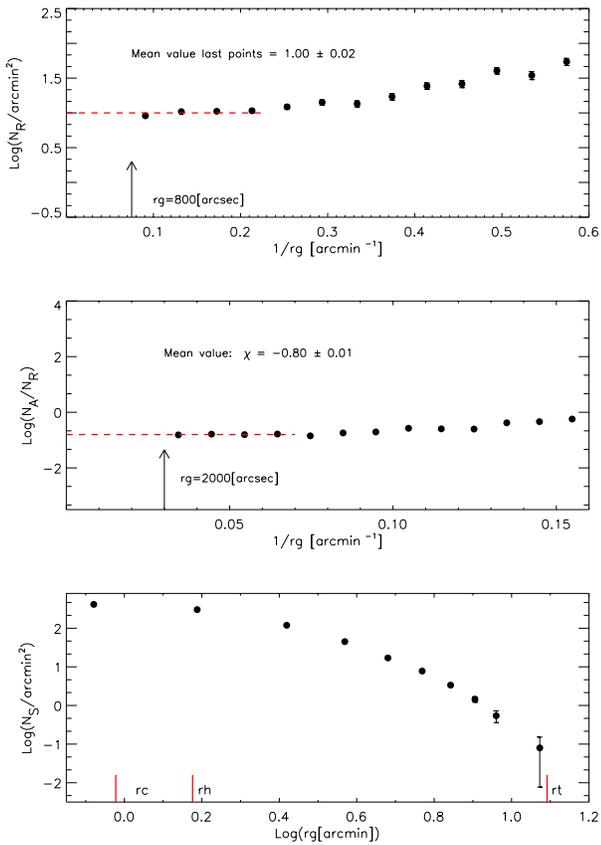}
\caption{{\it Top:} The logarithmic density of the candidate field stars is plotted
as a function of the inverse of the radial distance.{\it Middle:} The ratio between the number 
of accepted ($N_A$) and the number of rejected ($N_R$) stars, $\rm N_A/N_R$, 
as a function of the inverse of the radial distance.
{\it Bottom:} The density profile of the cluster 
with the number of spuriously accepted stars subtracted off.  Note the very different
ranges of radial distance shown in the three panels.}
\label{dense}
\end{figure}

The top panel of Figure~\ref{dense} shows the logarithmic density of
the number of ``rejected'' stars --[$N_R$]-- per arcmin$^2$) as a function of
the inverse of the radial distance.  True cluster members that happen to lie
within the rejection region of the CMD should represent a minor source of
contamination in this plot, becoming negligible at large distances (small
values of $\rm 1\over rg$).  The middle panel of the figure shows the number
ratio of stars in the box ($N_A$) to stars outside the box as a function of
the inverse radial distance.  Note that we can extend this curve to
greater radial distances from the cluster ($1\over r$ more closely approaching zero) than
in the upper panel because stars in the corners of the study area, outside
the fully sampled concentric annuli, can be used; the incomplete sampling at the
larger distances complicates the calculation of surface densities, but does
not affect the ratio of simple star counts.  We estimate that the log of the ratio of
field stars in the box to those outside the box approaches an asymptotic value
of $-$0.80$\pm$0.01.  The top panel of Fig. 4 indicates that the logarithm of the surface density of
{\it field stars\/} in the {\it rejection region\/} of the CMD is 1.0 $\pm$ 0.02.
Combined, these numbers yield a total surface density of 11.6 stars per square
arcminute:  10 stars per square arcminute fall outside the acceptance region in the
CMD, and 1.6 stars per square arcminute fall inside the box.

The bottom panel of Figure~\ref{dense} shows the density profile of the cluster
with the number of spuriously accepted stars subtracted off.  Because the radial
profile is not observed to flatten at the largest distances, we conclude that
the tidal radius of this cluster might be larger than previously estimated. 
Alternatively, we may be seeing evidence of a cluster halo, as has been
suggested for M92 \citep[by][]{testa00, lee03, dicecco09} and NGC~1851
\citep{ols09}.  In a forthcoming investigation, we plan to perform a more
detailed fit of the density profile to constrain the possible occurrence of a
halo around IC$\,$4499 and, in particular, whether the distribution of the
outermost stars is azimuthally symmetric.   


\section{Multiple Populations in IC$\,$4499}

It is now known that many GCs contain more than one stellar generation
(e.g., \citealt{pio09} for a review).  This manifests itself as abundance
anomalies on the red giant branch, such as the CH/CN dichotomy and the 
O/Na, Mg/Al anticorrelations, indicating that the stellar material has
been polluted with material released by a previous generation of
stars \citep{carretta10b}.

However, it is possible to detect the presence of multiple population via
$U$ band imaging \citep[e.g.][]{marino08,han09,kravtsov10a,kravtsov10b,lardo11}
as the CN and CH molecular features `line blanket' this bandpass and therefore
CH/CN strong giants will appear redder in a $U-(optical)$ color-magnitude
diagram. For example,\citet{yong08} showed that the Str\"{o}mgren $u$
traces the differences in N abundances for the stars of NGC 6752.
\citet{marino08} showed that the RGB stars in M$\,$4 have a bimodal
spread in a $U$ vs. $(U-B)$ CMD, where the red side of the RGB
is confirmed spectroscopically to be Na-poor and CN-weak and the
the bluer stars  are Na-rich and CN-strong.  \citet{carretta10b} showed
that Na abundance is correlated with the spread in $(\rm U-B)$ among RGB
stars in NGC 3201.  \citet{lardo11} showed that for the M5 stars, 
the Na-poor and Na-rich stars, which are tightly aligned along 
the narrow cluster RGB in the $g$ vs. $(g-r)$ CMD, are 
clearly separated into two parallel sequences in the much broader giant 
branch seen in the $g$ vs. $(u-g)$ diagram.  The Na-rich stars 
appear systematically redder than Na-poor ones.

These results show that $U$ photometry is a powerful 
diagnostic of light-element abundance spread; a spread
in the UV -- optical color on the RGB that is not seen in 
the optical -- optical color is evidence of light-element 
variation in RGB stars. Similarly, the DDO$\,$51 filter contains
numerous Mg features, including the prominent MgH indices,
and therefore scatter in this filter is sensitive to the Mg
abundance variations known to be a marker of the presence of
secondary stellar populations.  Here the $U$ band
and $DDO$\,$51$ band of the giant branch are used to check 
for possible signs of multiple populations in IC$\,$4499.

\begin{figure}
\includegraphics[width=9cm]{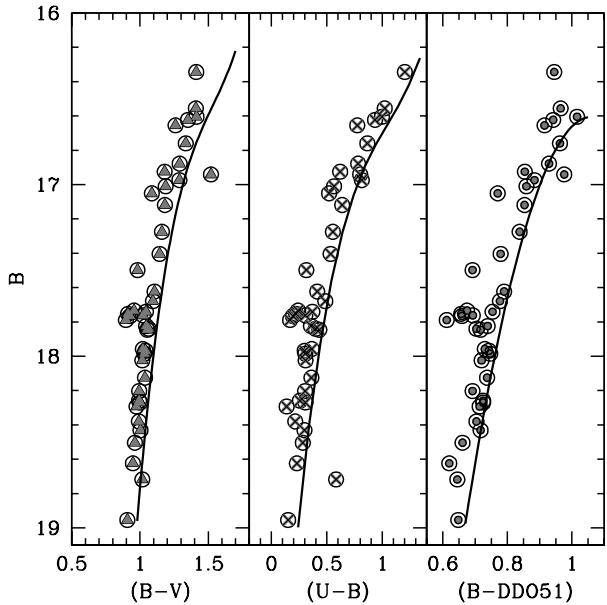}
\caption{The $B,(B-V)$, $B,(U-B)$ and $B,(B-DDO51)$ CMDs for 42 
giant stars in IC$\,$4499.  The curves approximately tracing the red edges of the 
RGBs are used as references to compute the color spread distributions shown 
in Figure~\ref{delCol2}.}
\label{BUDcmd}
\end{figure}

We will first consider the sample of giant branch stars from \citet{hankey10} in
a recent spectrographic survey of giants in a 2$^{\circ}$ field around IC$\,$4499.
Of the 43 stars found to be probable cluster members, 42 had good photometry in
our catalog.  Figure~\ref{BUDcmd} shows the \bb,\bmv, \bb,\umb, and \bb,\bmd \ CMDs of 
these stars.  Although this is a relatively small sample of stars, it has 
the advantage of being a decontaminated and unbiased giant sample. Following the 
\citet{lardo11} recipe for searching for bi-modality, a ridge line following the 
curvature of the observed RGB in \bb,\bmv, \bb,\umb, and \bb,\bmd \  and located 
approximately at the red edge of the RGB, is taken as a reference to compute
color spreads.  The ridge line adopted is a 4th order polynomial that closely 
follows the theoretical RGB from the BaSTI models.  


\begin{figure}
\includegraphics[width=9cm]{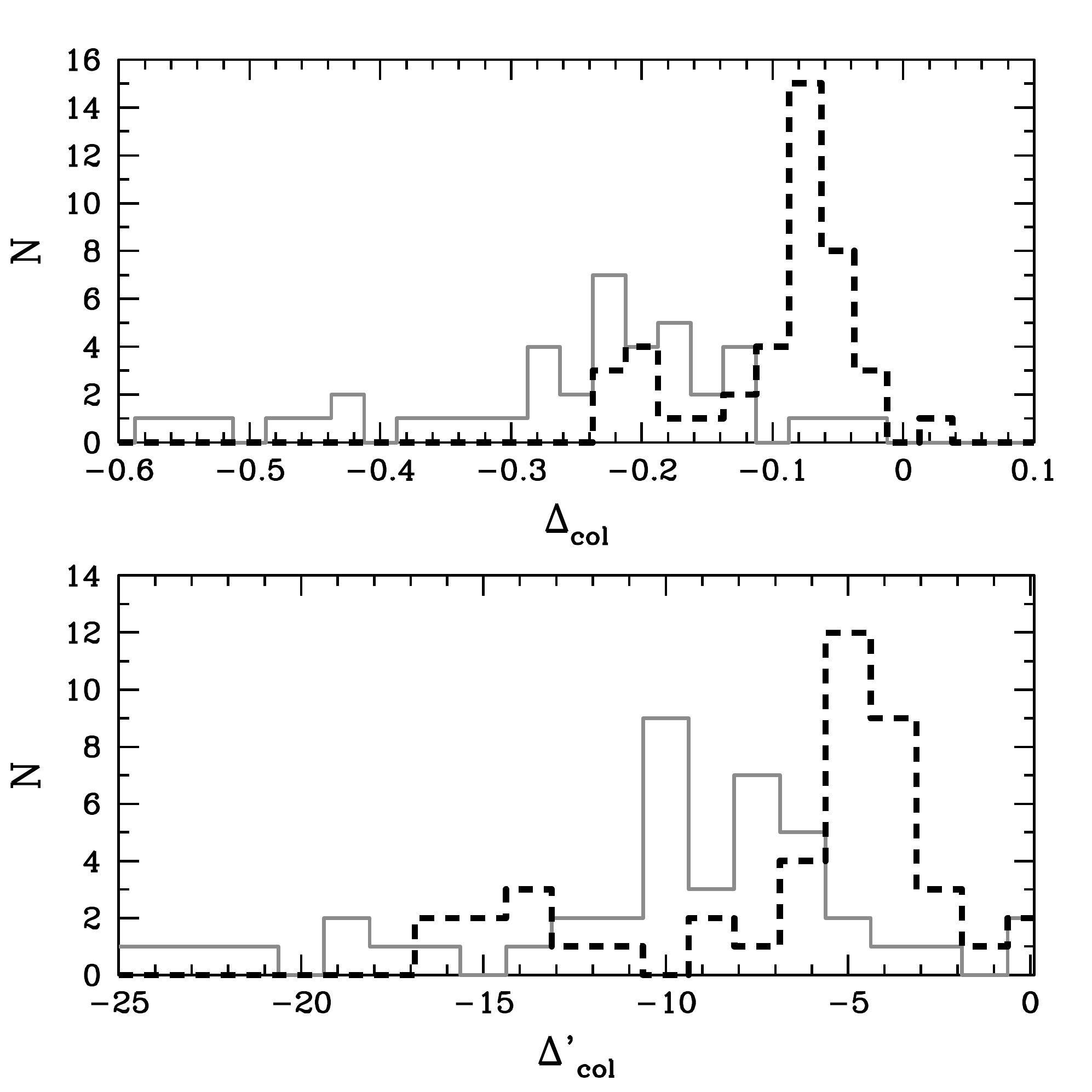}
\caption{{\it Top:} The color spread with respect to the RGB fiducials for the 
$(B-V)$ (dashed histogram) and $(U-B)$ (solid histogram) distributions.
{\it Bottom:} Same as above, but the color spread has been normalized to
take into account photometric errors.}
\label{delCol2}
\end{figure}

To take into account potential photometric errors that could affect
the color spread, the bottom panel of Figure~\ref{delCol2} shows the 
normalized color spread ($\Delta'_{col}$ = $\Delta_{col}$ / $\sigma_{col} $), the 
color spread of each star divided by the associated photometric uncertainty.
The photometric error estimates are based on the root-mean-square agreement 
of the actual observations.  There are 24 different telescope-CCD-filter combinations 
and dozens of observations per star.  Hence, the photometric errors will
be appropriate.

The wings of the $\Delta'_{col}$ distributions roughly coincide, and
the core of the \bmv \ distribution is more peaked than its \umb \ 
counterpart.  This was observed for the different GCs studies by 
\citet{lardo11}, but using the SDSS passbands and attributed to the 
spread in the abundance of light elements such as C, N, O, Na, etc.  

As shown by \citet{lardo11}, comparing $\Delta'_{col}$ distributions can reveal 
whether there is any significant spread in the \bb,\umb \ red giant branch in 
addition to that due to photometric errors.As the U band is sensitive to CN and 
CH molecular features in giant stars, a spread in \bb,\umb \ compared to \bb,\bmv 
\ can be attributed to variations in CN/CH line strengths in the giants.

This same process is repeated for $\Delta'_{\rm (B-DDO\,51)}$, to exploit the
sensitivity of the DDO51 filter to Mg indices.

To establish whether the \bmv and \umd or the \bmv and \bmd distributions are
different (and to what degree of significance), we carried out a Kolmogorov-Smirnov
test. Several iterations of the KS test were run and the value of the shift that maximizes 
the probability that the two samples are drawn from the same parent population, 
$P^{phot}_{KS}$, is found \citep{lardo11}.  This ensures that only the shape of 
the distributions between $\Delta'_{\rm (B-V)}$ and $\Delta'_{\rm (U-B)}$ is considered, 
and not potential unphysical shifts from, $e.g.$, the way the ridge line is defined 
(and hence $\Delta'_{col}$).  The $\Delta'_{\rm (B-V)}$ and $\Delta'_{\rm (U-B)}$ samples 
cannot be distinguished at a 93\% confidence level.   Similarly, the 
$\Delta'_{\rm (B-V)}$ and $\Delta'_{\rm (B-DDO\,51)}$ color spreads
cannot be distinguished at an 80\% confidence level.  These results are summarized 
in Table~\ref{tabP}.

\begin{figure}
\includegraphics[width=9cm]{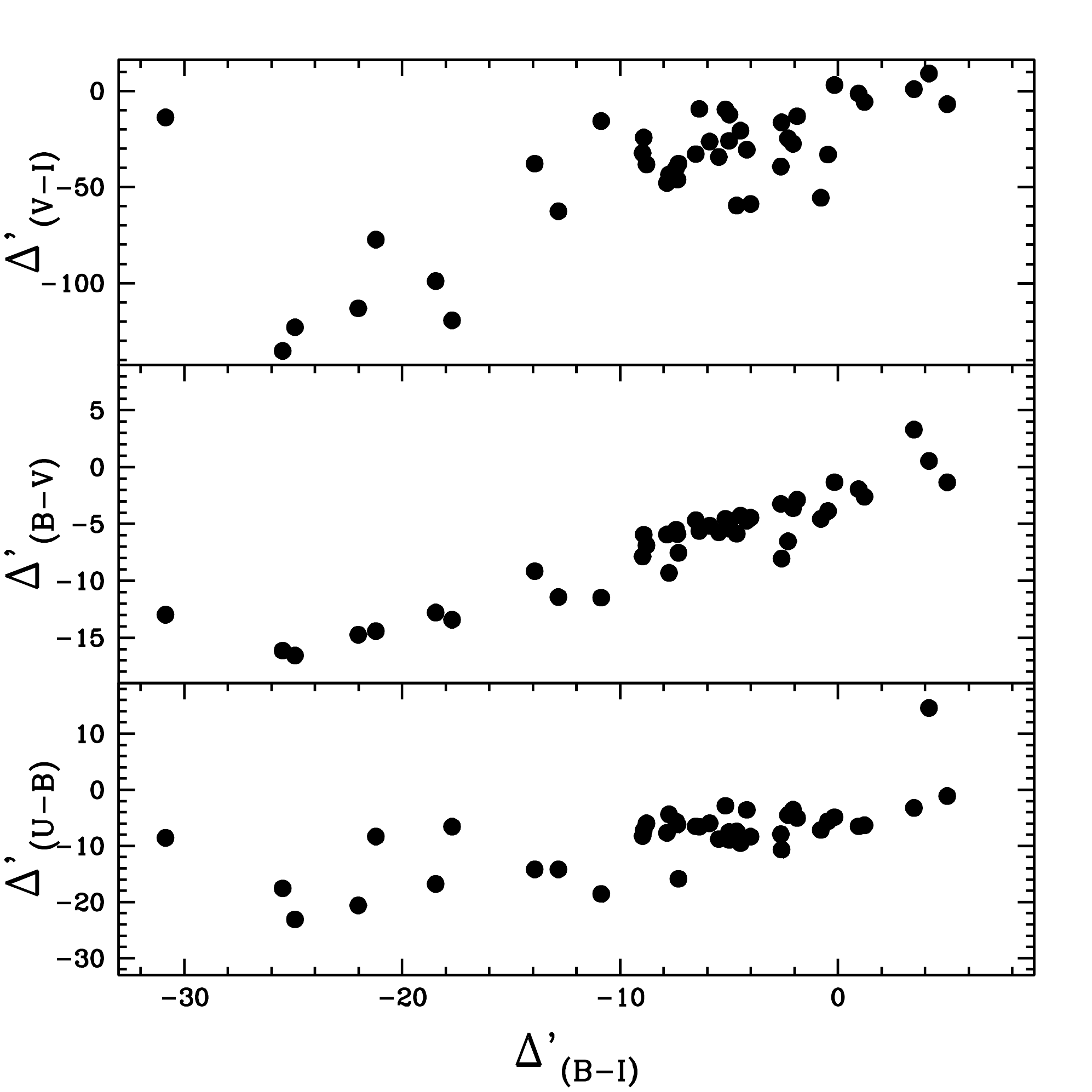}
\caption{The $\Delta'_{\rm (B-I)}$ versus 
$\Delta'_{\rm (U-B)}$, (bottom panel), $\Delta'_{\rm (B-V)}$,
(middle panel), and $\Delta'_{\rm (V-I)}$ (top panel).}
\label{PErrT3}
\end{figure}

The KS test indicates that the color distributions shown in Figure~\ref{delCol2} 
have the same shape.  However, it does not test whether the shapes 
are due to the same stars or completely different stars which have 
color distributions that produce a similar spread.  
Figure~\ref{PErrT3}
shows a positive, linear correlation for $\Delta'_{\rm (B-I)}$ versus 
$\Delta'_{\rm (U-B)}$, $\Delta'_{\rm (B-V)}$ and $\Delta'_{\rm (V-I)}$.  
This is strong evidence that the color-spread is intrinsic in
$\Delta'_{\rm col}$ and includes a significant component of temperature 
spread in not only the $(U-B)$, but also the $(B-V)$ and $(V-I)$.  Hence, the 
$\Delta'_{\rm (U-B)}$, $\Delta'_{\rm (B-V)}$ and $\Delta'_{\rm (V-I)}$
distributions all have a range of temperatures at fixed apparent magnitude.
This indicates that there is no spread in UV color on the RGB that is not 
seen in the optical.


\begin{table}
\centering
\caption{Results of KS tests}
\label{tabP}
\begin{tabular}{lcc}\hline
Color distribution & $\rm P^{phot}_{KS}$ & $\rm P^{phot}_{KS,all}$  \\
\\ \hline 
$\Delta'_{(\rm U-B)}$ & 0.93 & 0.58  \\
$\Delta'_{(\rm B-DDO51)}$ & 0.80 & 0.90 \\
\hline
\end{tabular}
\end{table}


\begin{figure}
\includegraphics[width=9cm]{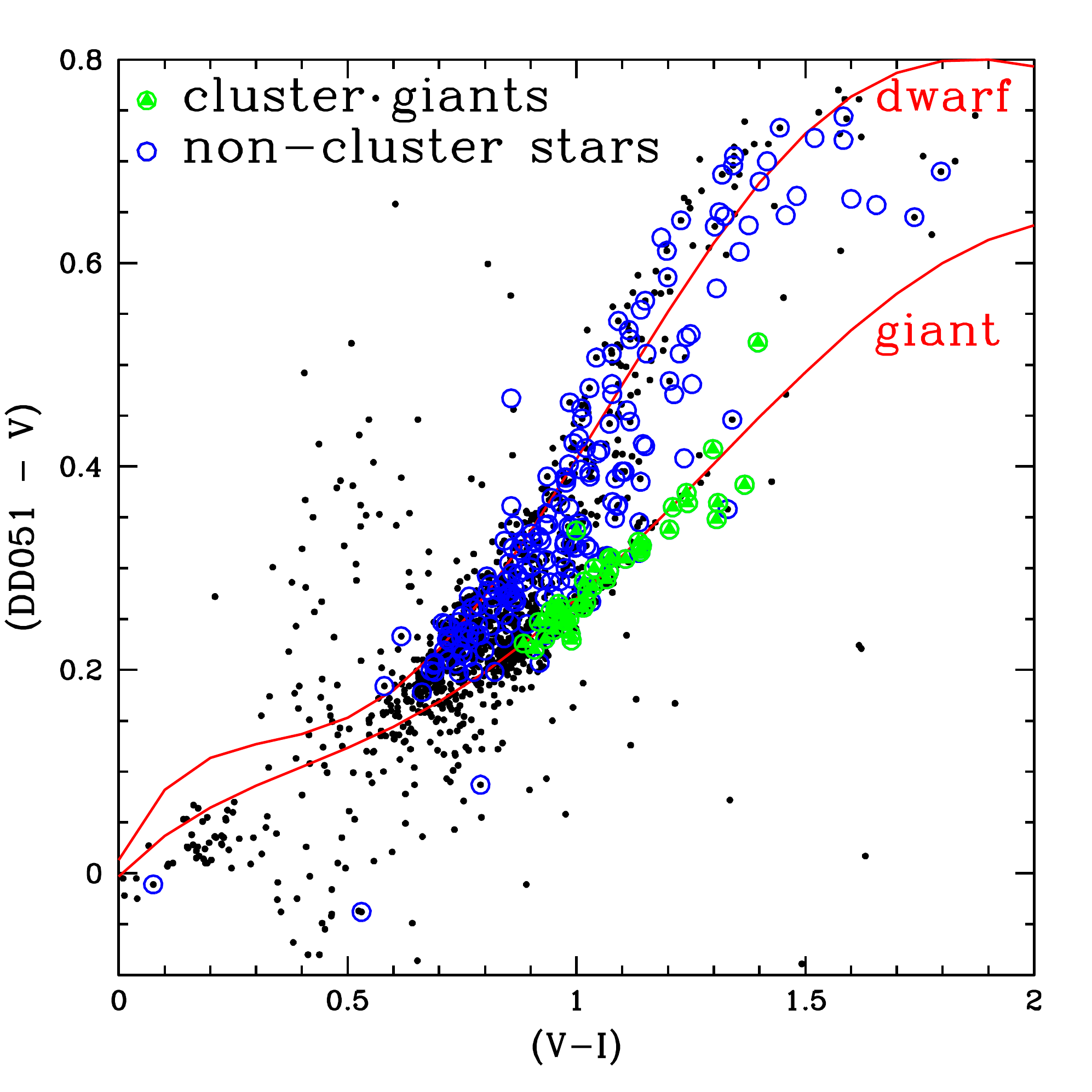}
\caption{The $(DDO51-V)$, $(V-I)$ color-color plot used to select giant
candidates in IC$\,$4499.  The green symbols represent cluster giants confirmed
spectroscopically by \citet{hankey10}, whereas the blue symbols represent stars 
that \citet{hankey10} found are not consistent with IC$\,$4499 cluster membership.
Polynomial fits used by \citet{teig08}
to separate M33 giants and dwarfs are also shown in red.}
\label{Dcmd}
\end{figure}

To increase the sample size, we will now consider giant stars selected using 
the DDO$\,$51 filter in a similar procedure outlined by \citet{teig08}.  Figure~\ref{Dcmd}
shows a color-color plot in \vmi, \ \dmv \ of all stars in the red giant 
magnitude range ($B$ $<$ 20 mag).  Over-plotted are
the confirmed cluster members from \citet{hankey10} as well as the stars
shown to have radial velocities and metallicities that are not consistent
with IC$\,$4499 membership.  The polynomial fits used by \citet{teig08}
to separate M33 giants and dwarfs are also shown, where the polynomial
is shifted to account for the reddening of the cluster.  Probable
giant candidates are chosen as stars within 0.01 mag of the polynomial
that represents the giants, and that have \dmv $>$ 0.9 mag.

Using these giant candidates, a KS test is run between the $\Delta'_{(\rm B-V)}$ 
and $\Delta'_{(\rm U-B)}$ distributions.  The two samples cannot be distinguished 
at a 58\% confidence level.   Between
the $\Delta'_{(\rm B-V)}$ and $\Delta'_{(\rm B-DDO\,51)}$ distributions,
$P^{phot}_{KS,all}$ is 90\%.  Again these results are detailed in Table~\ref{tabP}.

Using the $U$,$B$,$V$,$I$ and $DDO$\,$51$ passbands, 
no evidence for an anomalous color spread along the giant branch is seen 
for IC$\,$4499.  This result is in contrast to the more massive
GCs studied by \citet{lardo11}.  The photometry presented here not only has 
photometric uncertainties quite a bit smaller than the photometry used by
\citet{lardo11}--the $\Delta'_{col}$ distributions are $\sim$ 4 times more sensitive,
but also the multiple passbands allow for a greater variety of color spreads
to be detected.  Still, no anomalous color spread along the RGB is seen.  

If the anomalous color spread is due to variations in the abundances of light
elements, this effect would be weaker for
lower metallicities.  Hence, for GCs like IC$\,$4499, (clusters with \feh $>$ $-$1.7 dex)
this method may not be able to detect anomalous color spreads.  On the
other hand, \citet{lardo11} was able to detect an anomalous color spread
in M15 (\feh $\sim$ $-$2.26), M53 (\feh $\sim$ $-$1.99) and M92 (\feh $\sim$ $-$2.28),
which are substantially more metal-poor than IC$\,$4499.  

Therefore the absence of an anomalous color spread implies that any multiple 
stellar populations in IC$\,$4499 must have smaller differences in 
the light elements compared to more massive GCs in which color
spreads were detected using near ultraviolet bands.  This is in agreement
with the general idea that low-mass GCs do not have a deep potential well.
A deep potential well would allow the retention of ejecta of first generation stars 
and increase the chance of forming a second star generation \citep{dercole08}.  If this behavior is
found for other medium-mass GCs as well, then that would imply that lower-mass
GCs have simpler histories.

\section{Conclusions}
We present new $U$, $B$, $V$, 
$R$, $I$, and $DDO$\,$51$ photometry of the Galactic globular cluster
IC$\,$4499.  As the multiband photometry extends well below the main sequence 
turn-off, current BaSTI isochrones are fitted to three color baselines 
to derive an age of 12$\pm$1 Gyr.  This is the first modern age estimate for IC$\,$4499.
Despite previous suggestions of a 
younger age for IC$\,$4499 \citep{fer95}, we find that IC$\,$4499 is 
coeval with them the majority of Galactic GCs.  

The density profile of the cluster is measured to search for signs of tidal
streams and/or extra-tidal stars.  Out to $r\sim$800\sec, the density profile
continues to change.  This could be because the tidal radius 
is larger than previously estimated.  It could also be that  
IC$\,$4499 is surrounded by a halo, similar to that claimed for M92 
(Testa et~al. 2000; Lee et~al. 2003;  Di Cecco 2009) and in NGC~1851 
by Olszewski et~al. (2009).  More wide-field data is needed to constrain
the radial extent of IC$\,$4499.

A search for possible multiple components in IC$\,$4499 is carried out.
Using $U$, $B$, $V$, $I$, and $DDO$\,$51$, and
43 red giant branch stars confirmed spectroscopically by \citet{hankey10},
no anomalous color spreads potentially due to variations in 
the abundances of light elements are detected.   To
increase the sample size, the $DDO$\,$51$ narrow band filter
is used to select a sample of 100 probable giants.  Also with
this sample, no anomalous
color spread is detected.  This is in contrast to the more
massive GCs studied by \citet{lardo11}, in which 
different components using near ultraviolet bands were detected.
Given the small photometric errors in our catalog, this implies a
rather small spread of light elements among the giant branch stars
in IC$\,$4499. 

\section*{Acknowledgments}

This research has made use of the NASA/IPAC Extragalactic Database (NED) 
which is operated by the Jet Propulsion Laboratory, California Institute of Technology, 
under contract with the National Aeronautics and Space Administration. 
AdC, GB and SC have been partially supported by PRIN INAF 2009 (PI. Prof. R. Gratton).
MM is funded by the IAC (grant P3/94) and by the Science and Technology 
Ministry of the Kingdom of Spain (grant AYA2007-3E3507).


\label{lastpage}

\end{document}